# Video-speed Graphene Modulator Arrays for Terahertz Imaging Applications

Yury Malevich,[1,2] M. Said Ergoktas,[1,2] Gokhan Bakan,[1,2] Pietro Steiner[1,2], and Coskun Kocabas[1,2,3*]

1. Department of Materials, University of Manchester, Manchester, M13 9PL, UK
2. National Graphene Institute, University of Manchester, Manchester, M13 9PL, UK
3. Henry Royce Institute for Advanced Materials, University of Manchester, Manchester M13 9PL, UK

* Corresponding author email: coskun.kocabas@manchester.ac.uk

**Abstract:** Electrically tuneable high mobility charges on graphene yield an efficient electro-optical platform to control and manipulate terahertz (THz) waves. Real-world applications require a multiplex THz device with efficient modulation over a large active area. The trade-off between the efficient gating and switching speed, however, hinders the realization of these applications. Here, we demonstrate a large-format 256-pixel THz modulator which provides high-frame-rate reconfigurable transmission patterns. The time-domain and frequency-domain THz characterizations of graphene devices reveal the relaxation pathways of gate-induced charges and ion packing at graphene-electrolyte interface. The fundamental understanding of these limiting factors enables us to break the trade-off permitting switching frequencies up to 1 kHz. To show the promises of these devices, we demonstrate a single-pixel THz camera which allows spatial and spectroscopic imaging of large-area objects without any moving components. These results provide a significant advancement towards the achievement of non-invasive THz imaging systems using graphene-based platforms.

THz radiation promises a wide range of applications[1] including wireless communication,[2,3] non-invasive spectroscopic imaging,[4] biomedical diagnostics,[5] and product quality control[6] owing to the non-ionizing nature and submillimetre-wavelength. Being between optics and electronics, however, makes the THz waves challenging to access. Approaching the THz spectrum from the electronic side is limited by the switching speed of transistors. On the other hand, optical methods face the fundamental limitation due to thermal energy (1 THz ~ 4 meV, the thermal energy at room temperature ~ 26 meV). More importantly, the lack of an efficient electro-optic material for THz light has been hindering the development of THz optoelectronic components.

Over the last decade, various materials systems have been explored to control THz radiation; three approaches have outperformed the rest. The first approach relies on tuneable metamaterials where the loss is controlled by depletion of substrate charge carriers.[7] An electrode array using interconnected split-ring resonators fabricated on a doped epitaxial GaAs enables both intensity and phase modulation. Following this seminal work, metamaterial-based spatial light modulators (SLM) and imaging systems have been demonstrated.[8] The requirement of an epitaxial wafer, the complex microfabrication process and the narrow spectral response due the metamaterial resonance are the limitations of this method. The second approach uses photo-induced carriers generated on a semiconductor surface to control the transmission of THz light.[9] Illumination of silicon wafers with high power visible or near-infrared light photoexcites free charge carriers, thus alters the THz transmission. Structuring the illumination using micromirror arrays has enabled high-resolution THz imaging systems. The downside of this method is the need for high optical power to pump carriers in the Si wafer. The third approach relies on the electrostatic tuning of high mobility electrons on graphene, which yields Drude-like metallic behaviour at THz frequencies.[10,11] Altering the carrier concentration *via* electrostatic gating permits electrical control of reflection and transmission of THz radiation due to the tuneable optical conductivity.[12] The high carrier mobilities for both electrons and holes enable fast optical response covering the entire THz region. This method was first reported by Sensale-Rodriguez *et al.* using a back-gated graphene transistor[10] and attracted widespread attention. In their device, they used a graphene-dielectric-Si capacitor structure with the Si layer acting as a back-gate. Subsequently, this idea was extended to SLMs by combining back-gated devices into arrays.[13] However, the electrical breakdown of the dielectric layer limits the maximum charge density and consequently, the performance of these devices. Using plasmonic effects on patterned graphene or integrating metallic metamaterials improves the efficiency of the modulators.[11] Very recently, Chen *et al.* reported THz modulation by operating similar devices around the Brewster angle.[14] Steep variation of the reflectivity and phase around the Brewster angle enables significantly larger modulation of THz intensity and phase. Fabrication of these devices over a large active area is still challenging due to the requirement of an ultrathin large-area dielectric layer.

Electrolyte gating (EG) has been proposed as an alternative scheme to overcome the limitation of dielectric-based devices.[15–17] EG can provide an order of magnitude higher charge densities which can significantly alter the optical properties of graphene even in the visible wavelengths.[17,18] The critical enabling element of EG is the self-forming ultrathin electrical double layer (EDL) at the graphene-electrolyte interface that generates large electric fields (~$10^9$ V/m) in nanometre length scales without an electrical breakdown. The spontaneous formation of the EDL also eliminates the tedious thickness control, and yields charge accumulation over a large area without electrical short circuit. To exploit these superior features, we have been investigating electrolyte-based THz devices.[19–21] A trade-off between efficient gating and switching speed has always been a significant concern for practical applications. In this work, we break the trade-off by understanding the dynamics of the ion packing at the graphene electrolyte interface, which enables us to drive the modulators up to 1 kHz frequencies. We address the critical challenges associated with the electrolyte gating and device fabrication and demonstrate a large-format 256-pixel device with video-speed modulation. The results highlight the key capabilities and the scalability of our graphene devices without using any clean room processes or complex material preparation.

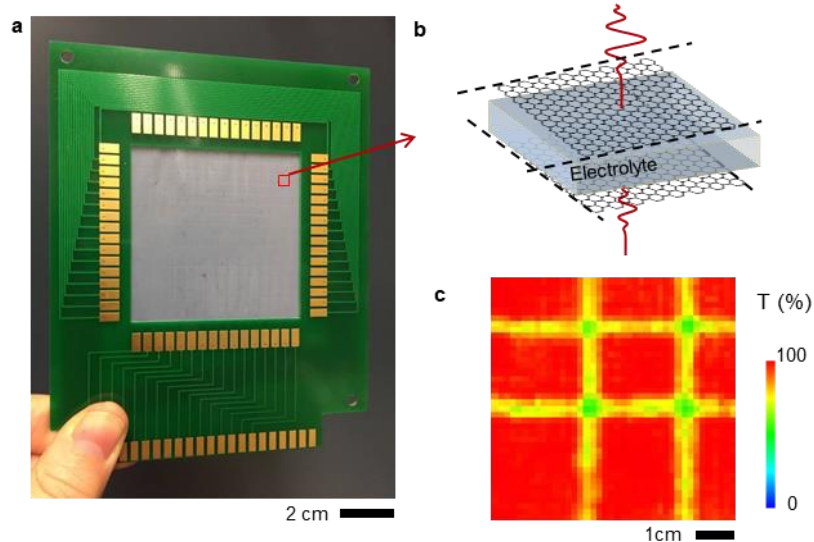

**Figure 1. THz spatial light modulator**. **a**, Photograph of the graphene-based THz spatial light modulator arrays (16×16) integrated on a printed circuit board. The local THz transmission is controlled by the passive matrix addressing of individual pixels. **b**, Schematic drawing of a single-pixel of the device consisting of 4-mm-wide graphene ribbons and electrolyte in between. **c**, A sample image of THz transmission (at 0.1 THz) through the device with two rows and columns biased at +1.0 and -1.0 V, respectively.

The fabrication process utilises the commercially available large-area chemical vapour deposition (CVD) grown graphene and the printed circuit board (PCB) technology. The device uses the passive cross-bar array technology with 16 columns and 16 rows (Figure 1a). The graphene stripes have resistance of 100 kΩ corresponding an intrinsic sheet resistance of 6.25 kΩ (See supplementary Figure S1). The stripes form the columns/rows acting as the active material and as well as the electrodes. Columns and rows are connected to individual contacts on a custom-design PCB. The intersection of each column and row defines a pixel (Figure 1b, see Materials and Methods section). A bias voltage applied between the corresponding row and column generates high mobility charges at the pixel. An external electronic circuit (National Instrument, NI 9264) controls the voltages on the rows/columns and creates a dynamic and spatially varying THz transmittance through the device. We use a room temperature THz camera (TeraSense-4096 based on an array of plasmonic detectors using high-mobility GaAs heterostructures) and a high power THz source (IMPATT diode with 100 mW output power at 0.1 THz) to image the transmittance of the device. Figure 1c shows THz transmittance image of the device with two rows and columns biased at 1.0 V and -1.0 V, respectively. After optimization of the device and driving electronics, we applied video-rate reconfigurable broadband transmittance pattern (see supplementary video 1).

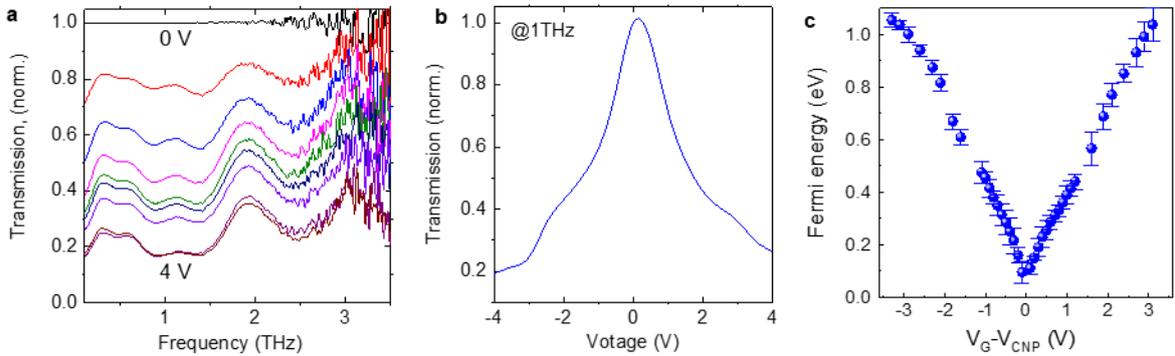

**Figure 2. Spectroscopic characterization of the graphene modulator**. **a**. Transmission spectra of the graphene modulator at bias voltages varying from 0 to 4 V with 0.5 V steps. **b**, Variation of the transmittance at 1 THz plotted against the bias voltage. **c**, Variation of the measured Fermi energy of gated graphene plotted against the effective gate voltage $V_G$-$V_{CNP}$ where $V_G$ is the gate voltage and, $V_{CNP} \approx 0.1$ V is the charge neutrality point. The Fermi energy is measured from the optical absorption spectra.

To reveal the high-performance capabilities of the device, we characterized a single-pixel device using time-domain THz spectrometer (Toptica TeraFlash). This system contains two fibre-coupled InGaAs antennas that can generate and detect THz pulses with > 4 THz bandwidth and over 90 dB dynamic range (Figure 2a). The device shows an efficient modulation (>70%) for the entire working range of the spectrometer. Fabry-Perot interference through the device yields variation of the THz modulation. Figure 2b shows the change of the normalized transmittance at 1 THz as the voltage bias varies between -4 to 4 V. The THz modulation is due to the Drude response of high mobility electrons and holes which yields frequency ($\omega$) dependent optical conductivity as $\sigma(\omega) = \sigma_{DC}/(1 - i\omega\tau)$ where $\sigma_{DC}$ is the low frequency electrical conductivity, $\tau$ is the electron scattering time which decays from 100 fs down to 50 fs as the Fermi energy changes from 0.1 to 1.1 eV.[22] The device shows the maximum THz transmittance at the Dirac point (~ 0.1 V). We observed 80% modulation for 1 THz at a bias voltage of 4 V. It is worth mentioning that, bias voltages > 3 V generate significant hysteresis and a slow response limiting the operation frequency of the device. To provide more insight into the ionic gating, we measured the Fermi energy as a function of bias voltage using optical absorption measurement (see supplementary Figure S4). Due to the Pauli blocking, the absorption of doped graphene shows an optical gap for energies < $2E_F$. The absorption spectra show a step-like change at the wavelengths corresponding to $2E_F$. The Fermi energy increases from 100 meV (at the charge neutrality point) to 1.0 eV at 3 V bias voltage (Figure 2c). The estimated charge density on graphene varies from $3.0 \times 10^{12}$ to $0.8 \times 10^{14}$ cm$^{-2}$.

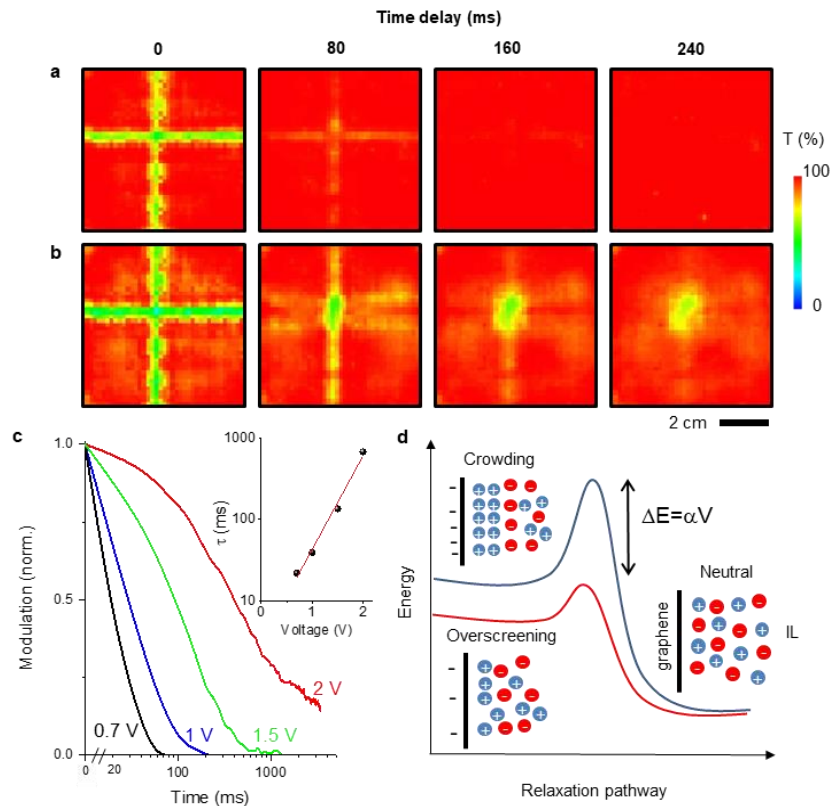

**Figure 3. THz imaging of carrier relaxation in graphene**. **a,b,** Snapshots of THz transmittance through the graphene modulator after a voltage pulse is applied between a row and column (1 V and 2 V used per one graphene layer, respectively). The relaxation time of the inducted charges varies exponentially with the bias voltage. **c,** Variation of the THz modulation showing the relaxation of the gate induced charges on graphene monitored by the THz transmittance for different bias voltages. (Inset) Relaxation time as a function of the voltage bias. **d**. Schematic graph showing the energy landscape during the relaxation process. The voltage-dependent energy barrier separates the bistable polarized and neutral states.

To achieve the high-speed modulation, we investigated the limiting factors underlying dynamics of charges on graphene. The intrinsic carrier relaxation time for pristine graphene is in the range of 0.1 to 3 ps.[23] For the electrolyte-gated graphene, we observed much longer decay times limited by the electrical double layer formation at the electrolyte-graphene interface. Understanding the dynamic polarization of room temperature ionic liquids is a challenge because they are composed of solely ions without any solvent that results in very long screening lengths. It is widely accepted that under a voltage bias, the interfacial profile consists of alternating cation- and anion-enriched layers[24] investigated by atomic force microscopy,[25–27] x-ray diffraction,[24,28,29] and surface forces apparatus.[30] For the first time, we use THz imaging to directly monitor the relaxation of gate-induced charges on graphene, which leads to the nature of the polarized electrolyte interface. We started our investigation by applying 1 ms voltage pulses to a single row and column of the modulator and monitored the THz transition at 50 frames/sec using a THz camera. Figure 3a and 3b show the snapshots of the THz transmittance recorded after 1 ms voltage pulses of 1 and 2 V, respectively. We observed that the relaxation time of the gate induced charges profoundly depends on the applied voltage (see supplementary video 2). Figure 3c shows the voltage dependence of the relaxation from the normalized THz transmittance. The relaxation time increases exponentially from 20 ms to 0.7 s as the bias voltage changes from 0.7 V to 2.0 V (inset in Figure 3c). This exponential dependence is the hallmark of an energy barrier limited relaxation process. The relaxation process requires reorganization of the cations and anions within the electrical double layers. Depending on the structure of the EDL, the coordinated motion of the ions experiences a bias dependent steric energy barrier as $\Delta E = \alpha V_b$ where $V_b$ is the bias voltage and $\alpha$ is the proportionality constant. The steric energy barrier[24] leads bistability of the layered structure of the anion/cation layers at the interface stabilized. We used the Néel relaxation theory to characterize the relaxation time of a bistable system as $\tau = \tau_0 \exp\left(\frac{\Delta E}{k_B T}\right)$ where $\tau_0$ is the intrinsic relaxation time without the barrier. The fitting of our data provides the steric energy barrier with a proportionality constant of $\alpha = 2.7\ ek_BT$. When the bias voltage varies from 0.7 to 2 V, the energy barrier increases from 47 to 135 meV.

We also observed that high bias voltages can also cause persistent charges on graphene with relaxation times longer than an hour. For more insight into the voltage-dependent packing of ions at the interface, we used a quantitative model developed by Bazant *et al.* using a Landau-Ginzburg-type continuum theory to explain the interplay between overscreening and crowding of ions at the interface[31]. At low bias voltages (~ 10$k_B$T/e ~ 0.26 V), strong correlations between the ions lead to overscreening where the first molecular layer at the interface contains more counter charge than the electrode surface. The charge neutrality is achieved after a few monolayers. The overscreening leads a relatively small steric energy barrier. High voltages (> 100$k_B$T/e ~ 2.6 V), however, lead to crowding of ions at the interface due to saturation of available lattice points. Thus the crowding extends across two monolayers with counter ions on the third monolayer. This extended crowding of ions at the interface introduces a substantial steric energy barrier resulting in very long relaxation time. Figure 3d illustrates the possible relaxation process which leads to the observed fast or slow decay constant depending on the bias voltage. Our results and the phenomenological model suggest that to achieve fast switching speed, crowding of ions at high voltages should be avoided.

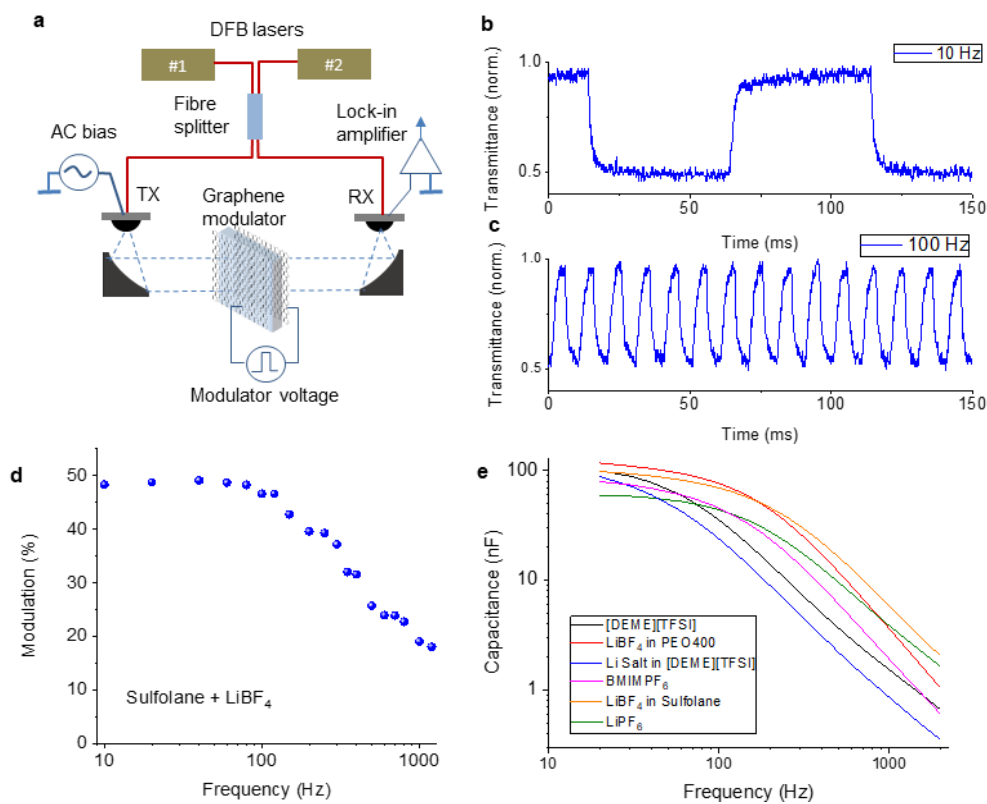

**Figure 4. The frequency response of a single pixel. a**, Experimental setup used for characterization of the frequency response of the single-pixel graphene modulator. CW terahertz radiation is generated using interference of two precisely tuneable 1550 nm DFB diode lasers on InGaAs photo mixes. A square wave voltage is applied to the modulator. **b,c** Time traces of the THz transmissions under square-wave voltages with 10 Hz and 100 Hz frequencies, respectively. **d,** Variation of the modulation amplitude as a function of bias frequency. **e,** Comparison of the frequency scaling of capacitance for various electrolytes.

While the camera setup provides large-scale imaging, its transient response is limited by its frame rate, preventing observation of high-speed dynamics of the gating process. To overcome this, we built a measurement system using continuous-wave (CW) spectrometer (Toptica Terascan) as shown in Figure 4a. The system generates CW terahertz radiation by optical heterodyning in high-bandwidth InGaAs photo mixer. The output of two CW-DFB lasers detuned around 1550 nm is converted into terahertz radiation, at the difference frequency of the lasers. An AC bias is applied to the transmitter and the signal is demodulated using a high bandwidth lock-in amplifier. The time-trace of THz transmittance through a 3×3 mm$^2$ single pixel device is recorded using an oscilloscope (Tektronix MDO3022). Figure 4b and 4c show time-traces under square wave excitations at 10 and 100 Hz. The device yields a flat 50% modulation up to an excitation frequency of 100 Hz. The device can operate up to 1 kHz frequency with a modulation amplitude of 15% (Figure 4d). Figure 4e shows the frequency-dependent capacitance of the devices measured with an LRC meter for six different electrolytes. The cut-off frequencies varying between 50 and 300 Hz. We obtained the best frequency response from a sulfolane-based electrolyte containing 1 M LiBF$_4$ salt. Although ionic liquid electrolytes show relatively slow responses, their low vapour pressure is an advantage for more stable operation and longer retention. Another critical aspect of electrolyte gating is the ability to operate at low bias voltages enabling integration of large number of devices with digital output of microcontrollers. We utilise a pulse-width modulation (PWM) scheme to control the transmittance of the modulator array. For our device, PWM has two main advantages: (1) It reduces the slow electrochemical effects on the graphene surface and (2) it is compatible with digital (TTL) outputs of microcontrollers which is a significant advantage for miniaturisation and integration of a large number of devices with the control electronics (see Supporting Materials Fig.S5).

To show the system-level integration, we demonstrate a THz imaging system based on a single detector and the graphene modulator array. Modern visible waveband cameras record the spatially resolved information using silicon-based focal plane array detectors placed at the image plane of the collection optics. A similar approach for THz light remains a challenge due to the lack of sensitive focal plane detector arrays and high power sources. Computational imaging using a modulator array and a single-pixel detector is a promising alternative[32]. This method relies on interrogating an object (*X*, the vector representing the transmittance of the object) with a series of spatially encoded patterns (*M*, the matrix containing all encoded patterns) and detecting the intensity of the scattered/transmitted light as $I = MX$. Then, computational imaging algorithms can reconstruct the image from the sequence of intensity measurements with the known spatially encoded patterns as $X = M^{-1}I$. Previously, reconfigurable THz masks were constructed using tuneable metamaterials[8] and structured illumination of silicon wafers.[9] In contrast, our THz modulator is a broadband electro-optic solution for THz imaging applications. Figure 5a illustrates the experimental setup. THz pulses are generated by the spectrometer, and collimated with a 3″ parabolic mirror. The THz modulator array is placed on the collimated optical path. To minimize diffraction, we place the object very close to the modulator and focus the transmitted pulses on the THz detector. Our time-domain spectroscopy (TDS) system provides fast acquisition of THz pulses (up to 50 pulses/s). We record the transmitted pulses while reconfiguring the mask. Figure 5b and 5c show three sample THz pulses and their spectra. The resolution of the spectrometer is sufficient to resolve the sharp absorption lines of water in the air. Figure 5d shows the variation of the pulse intensity when only a single pixel is modulated. The sequence of intensity measurements with known transmittance patterns enables us to reconstruct the image of an object. Figure 5e shows the reconstructed images of two large-area objects (5×5 cm$^2$) with cross and double-line patterns (photographs are shown in the insets) using the intensity measurement of 256 sequential masks. The passive matrix addressing of pixels limits the performance of the proof-of-concept THz imaging system. Higher performance for practical applications is expected with active-matrix addressing of an individual pixels that requires integration of backplane electronics.

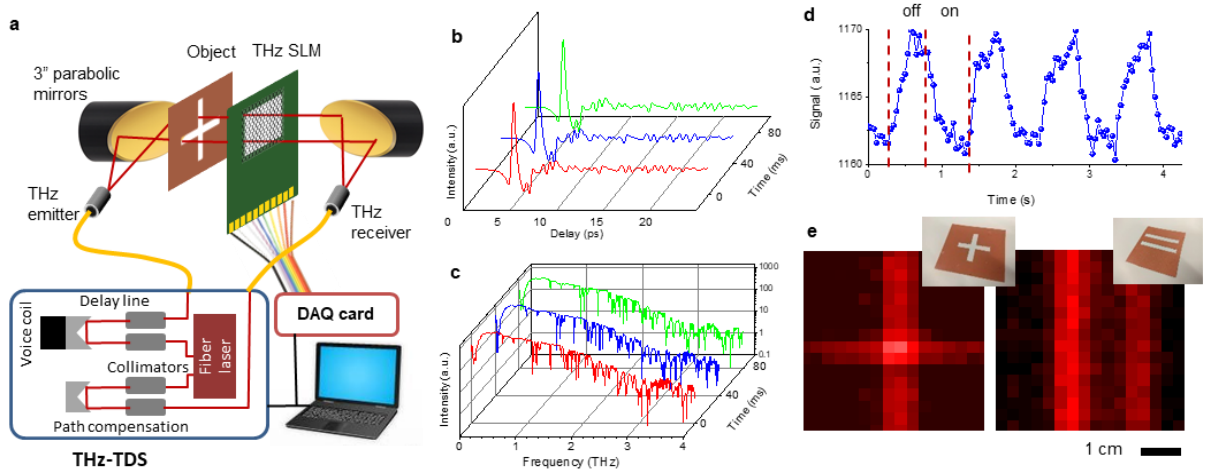

**Figure 5. Single-detector THz imaging system**. **a**, Schematic representation of the experimental setup consisting of a time-domain THz spectrometer, the graphene-based THz spatial light modulator, and data acquisition electronics. The object (metallic mask) is placed in front of the SLM on the collimated optical path. **b, c** Recorded THz pulses and their spectra acquired at a rate of 25 spectra per second. **d**, Variation of the integrated intensity between 0.1-2 THz while a single pixel of SLM is periodically switched on and off. **e**. Reconstructed images of two masks with cross and double-line patterns. Insets show the photograph of the metallic masks.

In summary, we demonstrated a graphene-based SLM that can generate reconfigurable THz patterns with video-speed frame rates. We presented the results of time-domain and frequency-domain THz characterization of the devices to understand the limiting factors underlying their limited switching speed. We observed that the voltage-dependent ion packing at the graphene-electrolyte interface is the key factor behind the dynamics of the electrolyte gating. By monitoring the relaxation of gate-induced charges, we were able to elucidate the steric energy barrier for the reorganization of cations and anions at the electrical double layer. This fundamental understanding of ion dynamics enables us to operate the THz modulator up to 1 kHz switching speed. To control the SLM with digital TTL signals, we implemented a PWM scheme that enables fast and hysteresis-free THz modulation. To show the promises of our device, we demonstrated a single-pixel THz imaging system using the developed THz modulator array as a reconfigurable transmission mask. We anticipate that the presented results provide a significant step towards the realization of THz imaging systems using graphene-based platform.

**Acknowledgements**: This research is supported by the European Research Council through ERC-Consolidator Grant (grant no 682723, SmartGraphene), and Defence Science and Technology Laboratory (DSTLX-1000135951).

**Author contributions**: Y.M., M.S.E., and C.K. designed and fabricated the devices. Y.M. performed the experiments. G.B. developed the PWM scheme and the control electronics. Y.M. and C.K. analysed the data and wrote the manuscript with feedback from the other authors. All authors discussed the results and contributed to the scientific interpretation as well as to the writing of the manuscript.

**Additional information**: Authors declare no competing financial interests.

**Materials and Methods**

**Terahertz measurements**: To investigate the frequency response of single-pixel devices we used Frequency-Domain Terahertz spectrometer (TeraScan from Toptica). It generates CW THz radiation by mixing light from two DFB diode lasers having precisely tuneable slightly different wavelengths near 1550 nm in an InGaAs photo mixer (Fig. 4a). A square wave voltage was applied to the modulator. The cut-off frequency of a modulator was determined by the LCR parameters of the circuit. The capacitance of single-pixel devices has been studied with Rohde&Schwarz Programmable LCR Bridge (Fig. 4e).

The experimental setup for the single-pixel imaging is shown in Fig. 6a. It is based on a time-domain THz spectrometer (TeraFlash from Toptica). THz radiation generated by the emitter passes through the object (metal mask), propagates through the SLM holding a certain transparency pattern, and detected by the detector. The pattern on the SLM was configured using National Instruments DAQ electronics.

**SLG transfer on PET**: Single layer graphene synthesized by CVD on copper foil was purchased from MCK Tech Co. Ltd. To fabricate the devices, we first laminated 75 μm-thick PET on single-layer graphene on Cu at 130 °C. Then, we etched the copper in 0.1M ammonium persulfate (APS) solution and rinsed the graphene with DI water after etching.

**Fabrication of spatial light modulator**: Two sheets of single-layer graphene on PET were patterned as 16 stripes of 4 mm width isolated from each other. 25 µm-thick porous polyethylene which performs as a separator and membrane was placed on one SLG sheet and ionic liquid electrolyte [DEME][TFSI] (Sigma Aldrich, 727679) applied afterwards. The other patterned SLG on PET was put on top as its graphene stripes are aligned perpendicular to the bottom ones. Each graphene stripe was connected to an individual electrode on the custom-design PCB using conductive carbon tape (Fig. 1a).

**Electrolyte Preparation:**

The following electrolytes used in this work are commercially available: [DEME][TFSI] (Diethylmethyl(2-methoxyethyl)ammonium bis(trifluoromethylsulfonyl)imide, Sigma Aldrich, #727679), BMIMPF$_6$ (1-Butyl-3 methylimidazolium hexafluorophosphate, Sigma Aldrich, #70956), and LiPF$_6$ in EC/DEC 1:1 (Lithium hexafluorophosphate in ethylene carbonate and diethyl carbonate from Gelon Energy Co., Ltd.). The other electrolytes prepared in house; 1M of LiBF$_4$ (Lithium tetrafluoroborate, Sigma Aldrich, #244767) in Sulfolane (Sigma Aldrich, #T22209), 1M of LiBF$_4$ in PEO400 (Polyethylene glycol 400, Alfa Aeasar, #B21992), and 1M Li Salt (Bis(trifluoromethane)sulfonimide lithium salt, Sigma Aldrich, #449504) in [DEME][TFSI].